# Spin and Orbital Angular Momenta of Light Reflected from a Cone

Masud Mansuripur*, Armis R. Zakharian[†], and Ewan M. Wright*

*College of Optical Sciences, The University of Arizona, Tucson, Arizona 85721
[†]Corning Incorporated, Science and Technology Division, Corning, New York 14831

**Abstract**. We examine several retro-reflecting optical elements, each involving two reflections. In the case of a hollow metallic cone having an apex angle of 90º, a circularly-polarized incident beam acquires, upon reflection, the opposite spin angular momentum. However, no angular momentum is transferred to the cone, because the reflected beam picks up an orbital angular momentum that is twice as large but opposite in direction to that of its spin. A 90º cone made of a transparent material in which the incident light suffers two total internal reflections before returning, may be designed to endow the retro-reflected beam with different mixtures of orbital and spin angular momenta. Under no circumstances, however, is it possible to transfer angular momentum from the light beam to the cone without either allowing absorption or breaking the axial symmetry of the cone. A simple example of broken symmetry is provided by a wedge-shaped metallic reflector having an apex angle of 90º, which picks up angular momentum upon reflecting a circularly-polarized incident beam.

**1. Introduction**. Electromagnetic (EM) waves carry energy, linear momentum, and angular momentum (AM). The rate-of-flow of energy (per unit area per unit time) at each point *r* in space and instant *t* in time is given by the Poynting vector $S(r,t) = E(r,t) \times H(r,t)$, where *E* and *H* are the local electric and magnetic fields, respectively. The linear momentum density is given by $S(r,t)/c^2$, where *c* is the vacuum speed of light. The angular momentum density with respect to an arbitrary point $r_o$ is $(r - r_o) \times S(r,t)/c^2$.

It turns out that two apparently different properties of EM waves can give rise to angular momentum: Circular polarization is one source of AM, which is usually referred to as the spin angular momentum (SAM). The other source is vorticity, which is associated with spiral phase variations around a given axis, and known as the orbital angular momentum (OAM) [1-6]. Experimentally, it is possible to distinguish the two types of AM, for example, by placing a small absorptive or birefringent particle in the path of the light beam. While the SAM of the beam causes the particle to spin on its own axis, the OAM will set the particle in orbital motion around the axis of the vortex [5, 7, 8].

The two types of AM are interchangeable, in the sense that a light beam can be made to interact with one or more optical elements in such a way that its SAM and OAM content (each as a fraction of the beam's total AM) will change as a result of interaction with the optical element(s). For example, a circularly-polarized Gaussian beam, which contains SAM only, may be sharply focused through a high-numerical-aperture lens, producing a focused spot that contains both SAM and OAM [9-11]. Alternatively, a circularly-polarized Gaussian beam may be sent through a specially-designed birefringent medium known as a "tuned *q*-plate," producing, upon transmission, a reversal of the sense of circular polarization in addition to an optical vortex of topological charge 2*q*, with *q* being an arbitrary integer [12]. While some of these processes may involve a net exchange of AM between the light beam and the optical element that is the catalyst for interconversion (e.g., a *q*-plate with $q \neq 1$), others will preserve the total AM of the light beam while converting a significant fraction of its SAM to OAM, or vice versa. A major goal of the present paper is to show the relative ease and flexibility of such interconversions with the aid of simple optical elements such as a hollow metallic cone or a solid dielectric cone.

In preparation for the analysis of reflection from a cone, we describe in Sec. 2 the properties of circularly-polarized light reflected from a flat plane, followed by an analysis in Sec. 3 of reflection from a 90º metallic wedge. In Sec. 4 we show how the reflection of a circularly-polarized Gaussian beam from a hollow metallic cone not only reverses the direction of the beam's SAM, but also endows the reflected beam with twice as much OAM. The case of solid dielectric cones involving two total internal reflections will be the subject of Sec. 7. We mention in passing that, although conical reflectors have been studied in the past for their applications in atom traps [13, 14] and as end-reflectors in certain types of lasers [15, 16], their ability to convert optical SAM to OAM (or vice versa) does not appear to have been noticed.

The study of angular momentum interconversion via reflection from conical mirrors reveals an interesting characteristic of axisymmetric objects. It turns out that an axially symmetric perfect electrical conductor (PEC) cannot possibly pick up any AM along its axis of symmetry, irrespective of the properties of the EM field that illuminates it. This result, which was first proved by Konz and Benford [17, 18], will be briefly described in Sec. 5. Although PECs provide an excellent approximation to good electrical conductors in the microwave regime, their usefulness as models for metallic objects in the optical regime is rather limited. For example, metallic particles typically used in optical trapping experiments generally absorb (within their skin depth) a fraction of the incident light, which is sufficient to transfer some optical SAM to these particles and, thereby, set them spinning. Thus, at optical wavelengths, it is perhaps more interesting to examine transparent dielectrics for their ability to pick up AM from an incident light field – without absorbing any photons, of course. It has been argued on general grounds that axisymmetric transparent dielectrics, much like PEC objects, cannot acquire angular momentum from incident EM waves [19]. The proof of this assertion, however, is not quite as straight-forward as that given for PEC objects by Konz and Benford [17]; it involves an expansion of the incident and scattered waves into their eigenmodes in spherical coordinates, followed by a demonstration that incident and scattered modes with differing azimuthal mode numbers do not couple to each other. In the special case where the *intensity* distribution inside a transparent axisymmetric dielectric exhibits symmetry around the same axis, the proof of the impossibility of AM transfer to the object is rather simple; this subject will be treated briefly in Sec. 8.

In light of the fact that SAM and OAM of EM waves can be rather easily transformed into one another, we will address in Sec. 9 the similarities and differences between these two types of optical AM, and probe the reasons behind their differing manifestations in experimental settings.

All the numerical simulations reported in the following sections are based on the Finite Difference Time Domain (FDTD) method [20]. In these simulations, the incident beam is started at $t=0$ in the source plane, which is typically an *xy*-plane located just above the object of interest. The beam is then propagated downward to interact with the object. The reflected beam, which returns in the general direction of the positive *z*-axis, is monitored once it reaches beyond the source plane.

**2. Reflection of circularly-polarized Gaussian beam from a flat reflector**. Consider a circularly-polarized Gaussian beam of light propagating along the negative *z*-axis, as shown in Fig. 1. For concreteness, we have assumed a beam with a vacuum wavelength of $\lambda_o = 0.5\,\mu$m, having a full-width-at-half-maximum amplitude (FWHM) of $4.0\,\mu$m. The amplitude and phase profiles of the three *E*-field components in a cross-sectional plane are shown in the figure. Also shown are the distributions of the *z*-component of the Poynting vector, $S_z$, as well as its transverse component $\boldsymbol{S}_\perp = S_x \hat{\boldsymbol{x}} + S_y \hat{\boldsymbol{y}}$. For the right-circularly-polarized (RCP) beam depicted here, the clockwise circulation of $S_\perp$ in the $xy$-plane gives rise to an AM aligned with the negative $z$-direction. The AM per unit-distance along $z$ is related to $S_\perp(x,y)$, to the vacuum speed of light $c$, and to the position vector $\boldsymbol{r} = x\hat{\boldsymbol{x}} + y\hat{\boldsymbol{y}}$, as follows:

$$J_z \hat{\boldsymbol{z}} = (1/c^2) \iint \boldsymbol{r} \times \boldsymbol{S}_\perp(x,y)\,\mathrm{d}x\mathrm{d}y. \tag{1}$$

For the beam depicted in Fig. 1, since $J_z$ is associated with the polarization state, it is commonly referred to as the beam's spin angular momentum (SAM). Note that $E_z$, which is substantially weaker than $E_x$ and $E_y$, exhibits a $2\pi$ vorticity in its phase profile; any orbital AM associated with this vorticity, however, should be negligible in the present example.

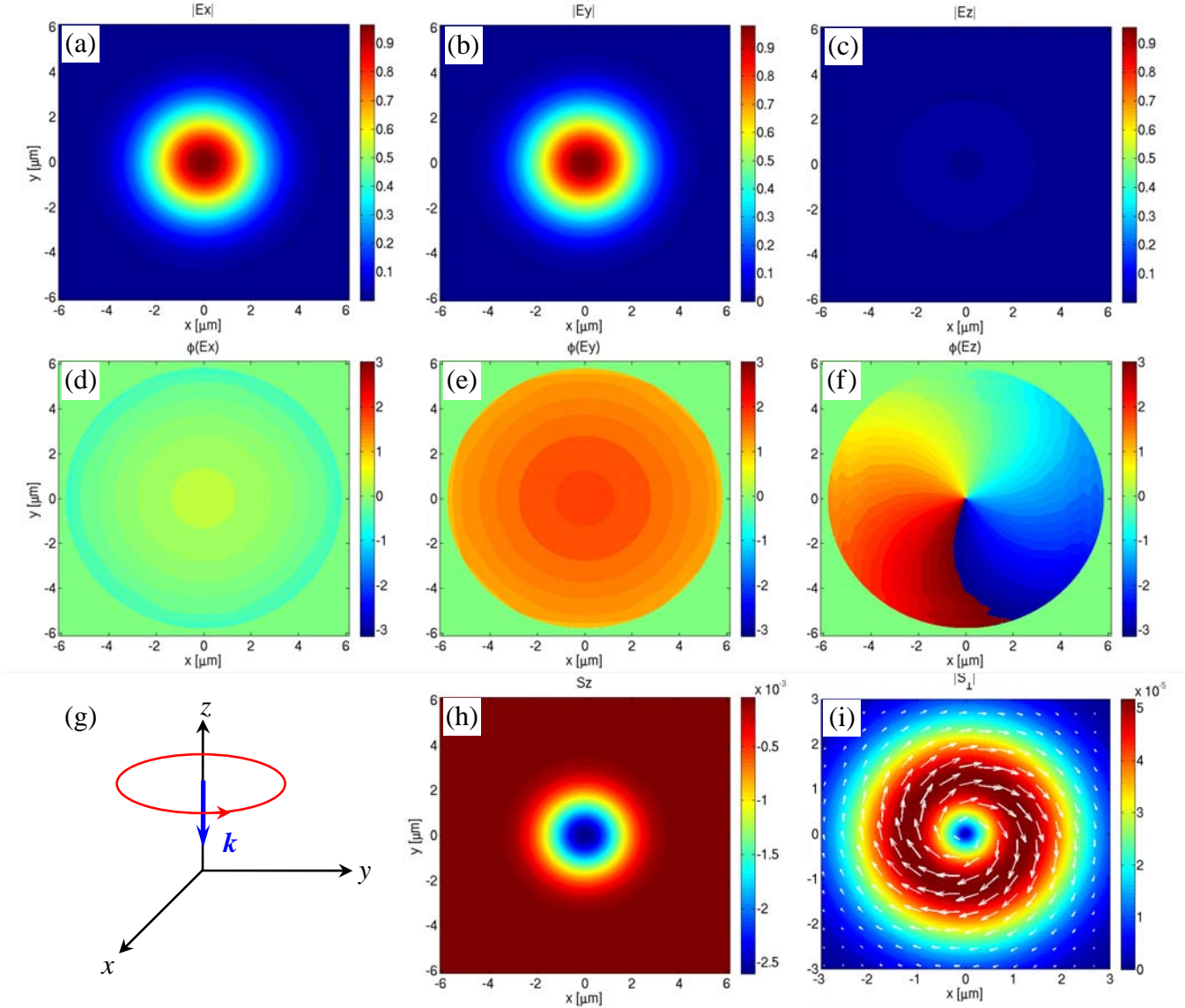

**Fig. 1** (Color online). RCP Gaussian beam propagating along the negative $z$-axis; $\lambda_o = 0.5\,\mu\mathrm{m}$; FWHM = 4 μm. Shown in the cross-sectional $xy$-plane are the amplitude and phase profiles of $E_x$, $E_y$, $E_z$, and also the Poynting vector components $S_z$ and $S_\perp$. The phase distributions of $E_x$ and $E_y$ are uniform: $\phi(E_x) = 0°$, $\phi(E_y) = 90°$. $E_z$ is relatively weak, but its phase profile shows a $2\pi$ vorticity around the $z$-axis. The clockwise circulation of $S_\perp$ around $z$ is responsible for the SAM of the beam.

Suppose now that the above beam is reflected from a flat PEC mirror in the *xy*-plane. At normal incidence, both $E_x$ and $E_y$ undergo a 180° phase-shift upon reflection, leaving the sense of rotation of the *E*-field unchanged. Of course the propagation direction is reversed, causing the reflected beam to be left-circularly polarized (LCP), but the direction of the angular momentum $J_z$ remains along the negative *z*-axis. This unchanging *z*-component of the beam's angular momentum upon reflection at normal-incidence implies that the mirror does *not* acquire an angular momentum in the process. In other words, the mirror does *not* tend to rotate in response to the incident beam – even though it picks up a linear momentum along the negative *z*-axis in consequence of the reversal of the direction of the light beam's linear momentum.

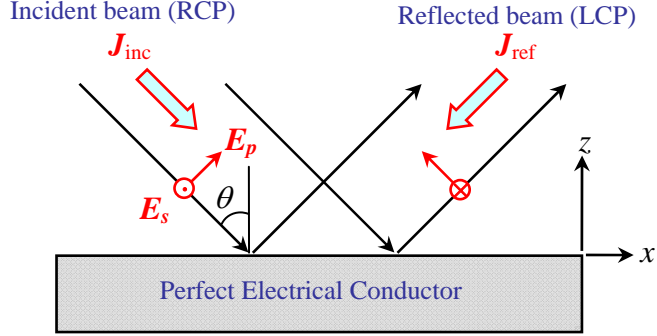

**Fig. 2** (Color online). RCP Gaussian beam arrives at the surface of a PEC mirror at the oblique incidence angle $\theta$; the incident AM is denoted by $\boldsymbol{J}_{\text{inc}}$. Upon reflection, the *s*-component of the incident *E*-field reverses direction, while the *p*-component reorients itself to remain perpendicular to the propagation direction. The projections on the mirror surface of both $\boldsymbol{E}_p$ and $\boldsymbol{E}_s$ of the incident and reflected beams cancel each other out at $z = 0$. The reflected beam is thus left-circularly-polarized, having the same AM content as the incident beam (i.e., $\boldsymbol{J}_{\text{ref}} = \boldsymbol{J}_{\text{inc}}$), but oriented differently. While the *z*-components of $\boldsymbol{J}_{\text{inc}}$ and $\boldsymbol{J}_{\text{ref}}$ are identical, their *x*-components are equal and opposite. This change of the *x*-component of AM upon reflection produces a net torque on the mirror, which tends to rotate it around the *x*-axis.

The above situation will change if the light is reflected from the mirror's surface at oblique incidence. As shown in Fig. 2, the incident beam's component of the *E*-field along the *s*-direction undergoes a 180° phase-shift upon reflection. In contrast, the *E*-field along the *p*-direction reorients itself in such a way as to remain perpendicular to the propagation direction, while also cancelling out the tangential component of the incident *E*-field at the mirror's surface. The net result is that the reflected beam, once again, becomes left-circularly polarized, with its SAM oriented opposite the beam's propagation direction. The change of angular momentum upon reflection is now given by $\boldsymbol{J}_{\text{inc}} - \boldsymbol{J}_{\text{ref}} = 2 J_{\text{inc}} \sin\theta \, \hat{\boldsymbol{x}}$, implying that the mirror must pick up this difference and begin to rotate around the *x*-axis. This result may indeed be confirmed by a direct calculation of the Lorentz force on the mirror's surface. The push of the *H*-field on the induced surface currents, and also the pull of the *E*-field on the induced surface charges, have slight asymmetries with respect to the *x*-axis, resulting in a net torque that causes the mirror to rotate around this axis. Reversing the sense of circular polarization of the incident beam will reverse the direction of the torque and, consequently, the direction of rotation of the mirror around *x*.

**3. Reflection from a metallic wedge**. Suppose now that the Gaussian beam depicted in Fig. 1 is reflected from a perfectly conducting 90° wedge, as shown in Fig. 3. The two successive bounces from the flat facets of the wedge cause the beam to return along the positive *z*-axis, albeit with the same sense of circular polarization as that of the incident beam (RCP in the present example). The change in the optical AM upon reflection is now given by $J_{ref} - J_{inc} = 2 J_{inc} \hat{z}$, causing the wedge-shaped retro-reflector to pick up this difference and begin to spin around *z*. The computed *E*-field amplitude and phase distributions of Fig. 4 confirm that the reflected beam is indeed RCP, and the plot of $S_\perp$ shows that the angular momentum of the reflected beam is opposite in direction to that of the incident beam.

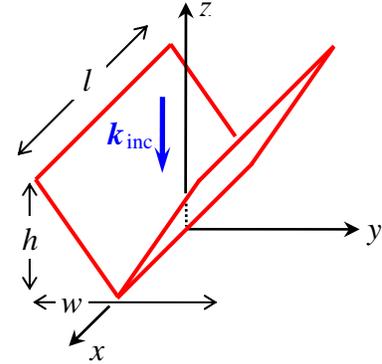

**Fig. 3** (Color online). Wedge-shaped reflector consists of two PEC sheets joined at 90°. The length, width, and height of the reflector used in the simulations are $l = w = 12 \, \mu m$, $h = 6 \, \mu m$. The RCP Gaussian beam of Fig. 1 is incident from above. After successive reflections at the sheets, the beam returns along the positive *z*-axis, having retained its RCP state of polarization.

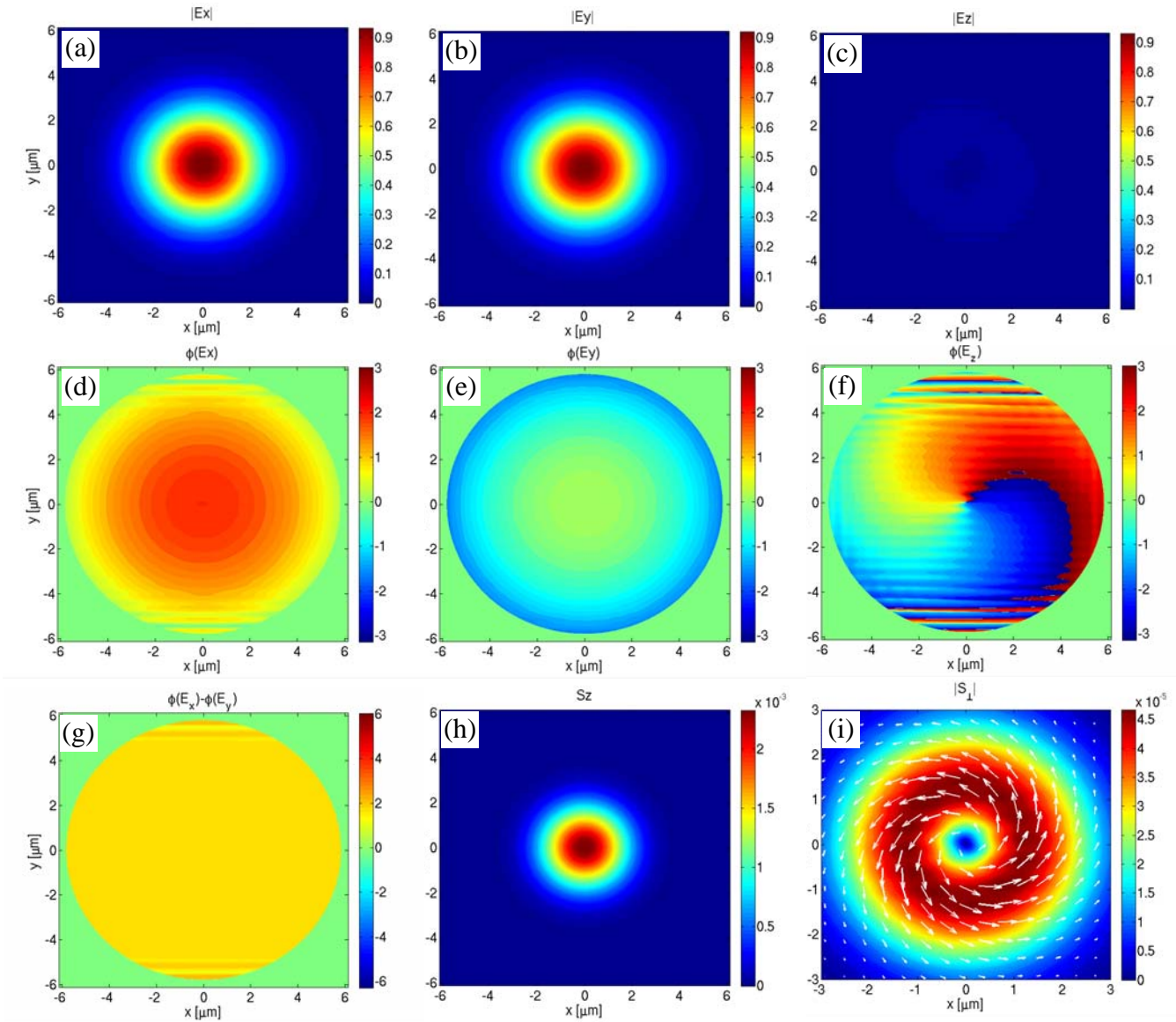

**Fig. 4** (Color online). Cross-sectional amplitude and phase profiles of $E_x$, $E_y$, $E_z$ for the Gaussian beam of Fig. 1, after reflection from the 90º PEC wedge depicted in Fig. 3. The phase profiles of $E_x$ and $E_y$ have acquired some curvature, but $\phi(E_x) - \phi(E_y) \approx 90º$. The $z$-component has retained its $2\pi$ vorticity, albeit with reversed handedness. The counterclockwise circulation of $S_\perp$ around $z$ confirms the reversal of the AM direction upon reflection from the wedge.

**4. Reflection from a metallic cone**. A naïve extension of the above ideas may lead one to believe that reflection of a circularly-polarized Gaussian beam from the PEC cone of 90º apex angle shown in Fig. 5(a) will result in the transfer of AM from the light beam to the conical reflector. This, however, is not what happens in reality. The simulation results shown in Fig. 6 reveal that the reflected beam no longer remains Gaussian, but becomes an optical vortex with a $4\pi$ phase winding. As expected, the SAM associated with the incident RCP beam reverses direction upon reflection. However, the beam also acquires twice as much OAM, associated with its vorticity, in such a way as to precisely cancel out the reversed SAM of the light beam. The end result is that the total AM of the reflected beam (i.e., spin + orbital) turns out to be exactly the same as the SAM of the incident beam. The absence of change in the light beam's AM, of course, means that the conical reflector will *not* pick up any rotational motion.

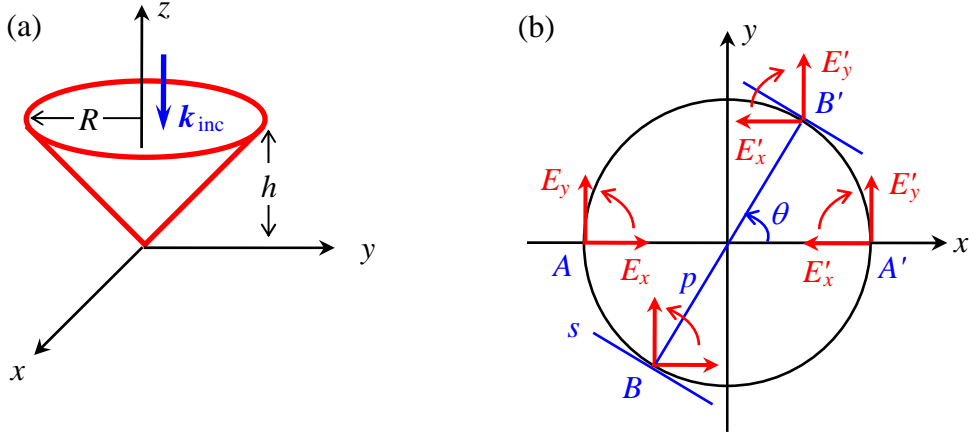

**Fig. 5** (Color online). (a) Hollow PEC conical reflector having a 90º apex angle. In our simulations, the cone's base radius and height were $R = h = 6$ µm. The RCP Gaussian beam of Fig. 1 is incident from above, along the negative $z$-axis. After two successive reflections at the conical surface, the beam returns along $+z$ with the same polarization state as the incident beam (RCP), but also having acquired vorticity with a $4\pi$ phase winding. (b) Cross-sectional view of the cone as seen from above. The ray arriving at time $t = 0$ at point $A$ on the cone surface is reflected twice before emerging at $A'$. For this ray, the $x$- and $y$-components of the $E$-field coincide with $p$- and $s$-polarization directions, respectively. Ignoring the constant time-delay needed for all such rays to traverse a given cross-sectional plane, the ray entering at $A$ will have the same phase as the ray emerging at $A'$. The two reflections are responsible for $E'_y$ being parallel to $E_y$, and $E'_x$ being anti-parallel to $E_x$. For the incident ray at $A$, the sense of polarization is right-circular, meaning that the $E$-field rotates from $E_x$ toward $E_y$. Similarly, the sense of rotation for the reflected ray emerging at $A'$ (also right-circular) is from $E'_x$ toward $E'_y$. With regard to the incident ray at $B$, since at $t = 0$ the $E$-field is aligned with the $x$-axis, it takes a certain fraction of the oscillation period for $E$ to arrive at the local $p$-direction. When this ray crosses the cone and arrives at $B'$, its $E$-field must rotate still further to come into alignment with $E'_x$. Therefore, the phase-difference between the rays emerging at $A'$ and $B'$ is twice the angle $\theta$ between the $x$-axis and the local $p$-direction at $B$. This explains the appearance of a $4\pi$ phase winding on the reflected beam that emerges from the cone.

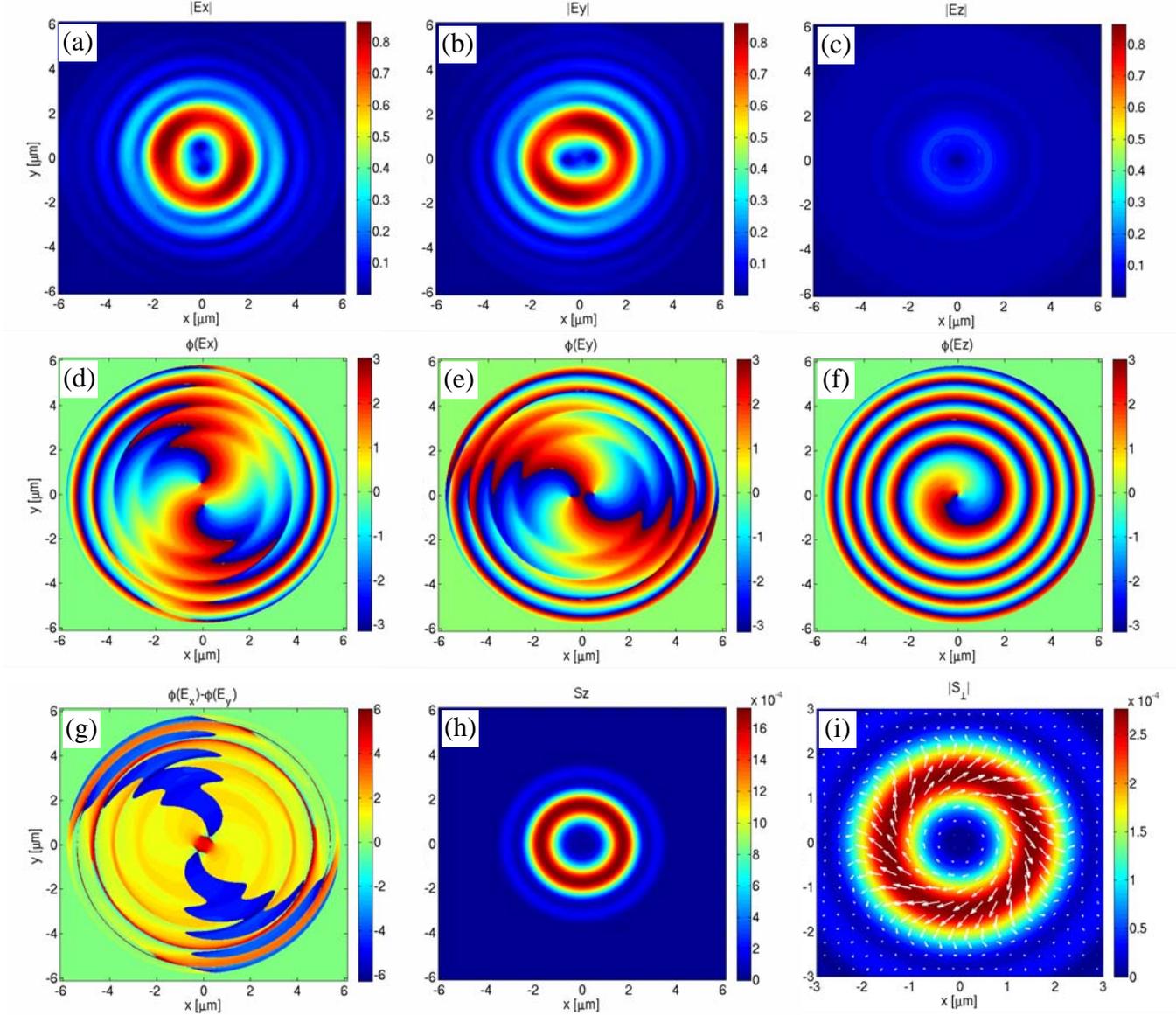

**Fig. 6** (Color online). Reflection of the RCP Gaussian beam of Fig. 1 from the hollow PEC cone depicted in Fig. 5(a). The reflected beam is also RCP, as can be seen in the plot of $\phi(E_x) - \phi(E_y)$, which is nearly constant at ~90° (modulo $2\pi$). Both $\phi(E_x)$ and $\phi(E_y)$ exhibit $4\pi$ vorticity around the $z$-axis; the phase of $E_z$, however, exhibits only a $2\pi$ spiral. The doughnut-shaped profiles of $E_x$, $E_y$, $S_z$ hint at the vorticity of the reflected beam. The clockwise circulation of $S_\perp$ around $z$ is dominated by vorticity, which opposes the counterclockwise circulation due to the beam's polarization state. The overall AM of the beam remains the same before and after reflection from the PEC cone.

A simple explanation for the appearance of OAM in the reflected beam may be given with the aid of the diagram of Fig. 5(b). In this top view of the conical reflector, the light ray striking the cone at point $A$ emerges at $A'$. At these locations, the $E_x$ and $E_y$ components of the $E$-field coincide with the $p$- and $s$-polarization directions associated with oblique incidence on the cone's surface ($p$ is perpendicular, while $s$ is tangential to the local surface). Ignoring the time-delay associated with the ray in its passage from $A$ to $A'$, we see that the RCP $E$-field of the incident ray at $A$ rotates from $E_x$ toward $E_y$, while the corresponding $E$-field of the (twice reflected) ray at *A'* rotates from $E'_x$ toward $E'_y$. Next, consider the ray incident at *B* and emergent at *B'*. Here the *p*- and *s*-polarization directions no longer coincide with the *x*- and *y*-axes. For the ray incident at *B*, it takes a fraction of the oscillation period for the *E*-field initially aligned with the *x*-axis to assume the *p*-orientation. Then, upon crossing the cone and arriving at *B'*, it must rotate still further before assuming the orientation of $E'_x$. It is these time-delays that determine the phase of the emergent ray at *B'* relative to that at *A'*, and are, therefore, responsible for the $4\pi$ phase winding imposed on the reflected beam. The twist in the *s-p* orientation around the cone is similar to the rotation of the birefringence axis in *q*-plates, albeit without reliance on anisotropy.

An uncanny similarity exists between the way a PEC cone reverses the spin of an optical beam while imprinting upon it a helical phase, and the way in which the adiabatic inversion of the bias magnetic field in an Ioffe-Pritchard trap flips the individual atomic spins of a Bose-Einstein condensate (BEC) while imparting vorticity to the condensate as a whole [21]. The helical phase thus imposed on the BEC can be interpreted as a Berry's phase [22]. By analogy, we may refer to the $4\pi$ phase spiral associated with the beam reflected from the PEC cone (and depicted in Fig. 6) as a Pancharatnam-Berry phase [23,24]. Incidentally, this terminology is also used in conjunction with the vorticity imparted to a circularly-polarized beam upon passage through a *q*-plate [12].

A second example from atomic physics is the coherent transfer of orbital angular momentum from an atomic system to a light field [25]. In the reported experiment, a rubidium vapor sample was given a spatially-varying quantum phase by using a spin degree of freedom. The $^{87}$Rb atoms acquired orbital angular momentum through Larmor precession in a magnetic quadrupole field. Subsequently, a Gaussian control beam propagating through the vapor cell generated a Laguerre-Gaussian beam that had the expected degree of vorticity.

The fact that the PEC cone does not absorb energy from the incident light, and also its circular symmetry around *z*, are sufficient grounds upon which to prove the impossibility of transferring optical AM from a light beam to the material object; a simple proof is presented in Sec. 5. It should therefore come as no surprise that, in the process of reflection from the hollow PEC cone, the incident beam, whose SAM is destined for reversal, goes through the trouble of acquiring the requisite amount of OAM in order to maintain its overall angular momentum.

We mention in passing that shifting the cone laterally by a short distance away from the center of the incident beam does not alter the main conclusions of the preceding analysis. In particular, when the cone was shifted by 2 μm along the *y*-axis, the reflected beam retained its total optical power and also its angular momentum along *z*, confirming once again that, in the absence of absorption, the cone cannot be made to spin around its own axis, even when the incident beam lacks symmetry with respect to that axis [17-19].

**5. Impossibility of optical AM transfer to perfect electrical conductors of axial symmetry**. We reproduce here the argument of Konz and Benford [18] in order to show that the PEC cone cannot possibly acquire angular momentum in its interactions with the EM field. Since EM waves cannot penetrate below the surface of PEC objects, the Lorentz force of the light can only act on charges and currents induced on the surface(s) of such media. The Lorentz force density is given by

$$\boldsymbol{F}(\boldsymbol{r},t) = \sigma(\boldsymbol{r},t)\boldsymbol{E}(\boldsymbol{r},t) + \boldsymbol{\mathcal{J}}_s(\boldsymbol{r},t) \times \mu_o \boldsymbol{H}(\boldsymbol{r},t). \qquad (2)$$

This equation describes the density of the EM force at point *r* of the surface at time *t*, produced by the action of the local *E*-field on the surface charge density $\sigma(\boldsymbol{r},t)$, and by the action of the local *H*-field on the surface current density $\mathcal{J}_s(\boldsymbol{r},t)$; here $\mu_\text{o}$ is the permeability of free space. Now, since the tangential *E*-field at the surface of a PEC must be zero, the local *E*-field everywhere is perpendicular to the surface. Also, since the component of the *H*-field perpendicular to the PEC surface is necessarily zero, both the local *H*-field and the surface current density $\boldsymbol{J}_s$ must be parallel to the surface at each and every point. Consequently, the force acting on any point of a PEC surface is perpendicular to that surface. Axial symmetry of the object around *z* now ensures the vanishing of the EM torque along the *z*-axis – because any force vector that is perpendicular to the surface is also co-planar with the *z*-axis. This completes the impossibility proof of transferring EM angular momentum to an axisymmetric PEC object along its axis of symmetry.

Note that the above proof is independent of the properties of the light beam; in particular, the beam is not limited by any symmetry constraints. Also, there is no prohibition of AM transfer to the object along directions other than those of its symmetry axes. Finally, metallic objects that are good (but not perfect) electrical conductors do not satisfy the requirements of the above theorem. This is because the tangential *E*-field at the surface, as well as the perpendicular *H*-field within the skin-depth, of such metallic objects are not exactly zero. To the extent that real metallic objects absorb energy from the EM field, they are capable of absorbing AM as well, and can, therefore, be expected to pick up rotational motion (even along their axes of symmetry) upon interacting with EM waves.

**6. Focusing circularly-polarized light via a parabolic reflector**. A variation on the theme of conical reflectors is provided by the PEC parabolic mirror shown in Fig. 7. Assuming the paraboloid's height above the *xy*-plane as a function of the radial distance *r* is given by $h(r) = a r^2$, the focal point of the mirror will be a distance of $1/(4a)$ above the *xy*-plane on the *z*-axis. Choosing the entrance aperture radius as $R = 6\,\mu\text{m}$, and setting $a = 1/12\,\mu\text{m}^{-1}$, we find the height of the entrance aperture as $h = 3\,\mu\text{m}$, and the focal point *F* of the mirror located at $z = 3\,\mu\text{m}$. The paraboloid focuses the collimated Gaussian beam of Fig. 1 into a diffraction-limited spot at its focal plane, as shown in Fig. 8. Note that, unlike the case of the hollow cone of Fig. 5, each incident ray is reflected only once from the paraboloidal surface.

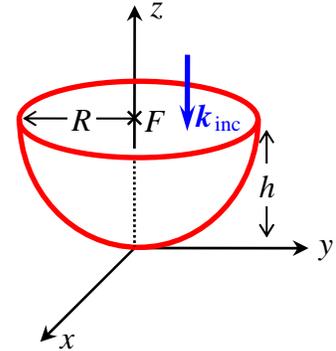

**Fig. 7** (Color online). Parabolic PEC mirror brings a collimated incident beam propagating along the negative *z*-axis to diffraction-limited focus at its focal point *F*. The height of the simulated mirror is $h = 3$ μm, its radius at the top is $R = 6$ μm, and its focal plane is located 3 μm above the *xy*-plane.

The theorem proven in the preceding section informs us that no AM should be transferred to the mirror and that, therefore, the incident and reflected angular momenta must be identical. The fact that the SAM of the incident rays is along the *z*-axis, whereas the rays reflected from the paraboloid have their SAM tilted away from *z*, implies that the SAM content of the beam must drop in the process of focusing. Conservation of optical AM would then require the conversion of some of the incident SAM to OAM. This is the same argument as has been made in the past with regard to focusing circularly-polarized light through an axially-symmetric lens [9-11]. We present here the example of a parabolic mirror not as a novelty in itself, but as a means to compare and contrast its SAM-to-OAM conversion mechanism with that of the PEC cone.

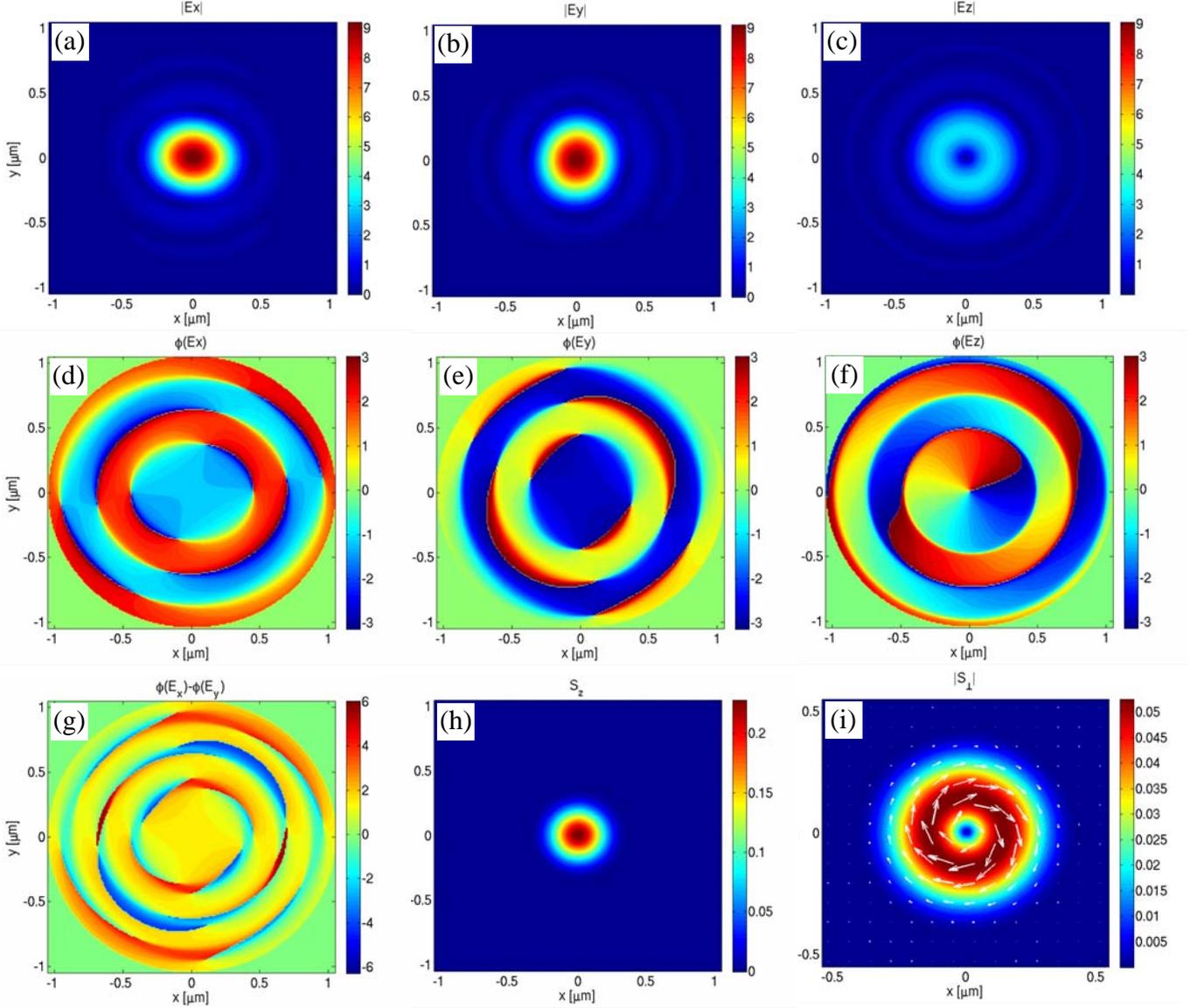

**Fig. 8** (Color online). Various properties of the reflected beam at the focal plane of the paraboloid depicted in Fig. 7; the incident beam is that shown in Fig. 1. Whereas $E_x$ and $E_y$ are vortex-free, there is a $2\pi$ vorticity in the $E_z$ component of the reflected beam. While the transverse component $S_\perp$ of the Poynting vector contains contributions from both SAM and OAM, it is difficult to isolate their individual contributions in these pictures. The integral of $S_z$ over the $xy$-plane yields the total optical power of the reflected beam, which turns out to be equal to that of the incident beam, as expected. The integral of $\boldsymbol{r} \times \boldsymbol{S}_\perp / c^2$ over the focal plane, however, overestimates the total AM of the beam by as much as 44%; this is due to the substantial departure from paraxiality of the focused beam.

With reference to the various plots of Fig. 8, we note that the AM of the focused beam, which is associated with the net circulation of $\boldsymbol{S}_\perp$ around $z$, is a complex mixture of spin and orbital components, although it is difficult to pinpoint the individual contributions of SAM and OAM to the total AM. While the vorticity of the focused beam is entirely in the $z$-component of the $E$-field, that of the beam reflected from the PEC cone (and depicted in Fig. 6) is primarily associated with the $E_x$ and $E_y$ components. Clearly, the SAM to OAM conversion must occur at the surface of the parabolic mirror, even though none of the light beam's energy or AM is transferred to this reflector.

Because of departure from paraxiality, the AM of the beam reflected at the parabolic mirror could no longer be computed by integrating the angular momentum density in a single cross-sectional plane of the beam; rather, it was necessary to illuminate the reflector with a short pulse of light, then monitor the entire AM of the pulse before and after reflection. When this procedure was followed, we found that the total AM of the wavepacket was conserved.

If the parabolic mirror of Fig. 7 were to be extended beyond its current height of $h = 3\,\mu\text{m}$, the spreading of the focused spot (while propagating along $+z$) would be arrested by the extended paraboloid, the reflected beam becoming collimated once all its rays bounce off the (extended) mirror for a second time. The emerging (collimated) beam would then have a reversed SAM relative to the incident beam, and also an acquired $4\pi$ vorticity, similar to that of the beam reflected from the PEC cone shown in Fig. 6. Seen from this perspective, the focused spot of Fig. 8 occupies a "halfway house" between the incident beam of Fig. 1 and the fully-converted beam of Fig. 6.

**7. Reflection of circularly-polarized light from a transparent dielectric cone**. In place of the hollow PEC cone described in Sec. 4, we would now like to employ a solid dielectric cone to retro-reflect the Gaussian beam depicted in Fig. 1. A first requirement for the dielectric cone is that its refractive index $n_{\text{cone}}$ be large enough that the 45° incidence angle on its conical surface exceed the critical angle of total internal reflection (TIR). A second requirement is for the cone to have an anti-reflection coating on its top surface to ensure that the entire beam enters the cone and then, following two internal reflections, leaves the cone. (For numerical computations, the simplest anti-reflection coating is a quarter-wave-thick dielectric layer of refractive index $\sqrt{n_{\text{cone}}}$; see Fig. 9.) Even after the above considerations, the dielectric cone differs from the PEC cone in one important respect: the phase difference between the reflected $E_p$ and $E_s$ components of polarization depends on the cone's refractive index, $n_{\text{cone}}$, thus introducing different states of elliptical polarization in the reflected beam.

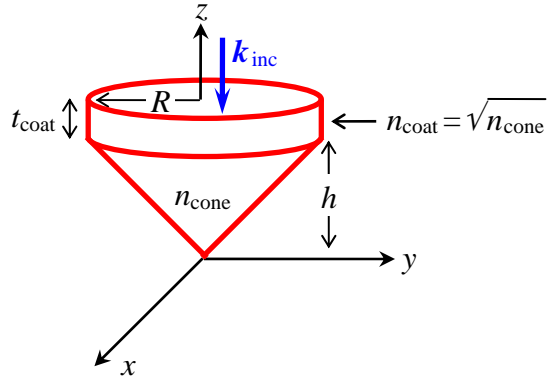

**Fig. 9** (Color online). Transparent dielectric cone having a 90° apex angle and an antireflection coating layer on the top facet. In our simulations, the base radius and the height were $R = h = 6\,\mu\text{m}$, the refractive index of the coating layer was the square root of the cone's index, and the coating layer's thickness was $\lambda_0/(4n_{\text{coat}})$. The RCP Gaussian beam of Fig. 1 is incident from above, along the negative $z$-axis. The beam returns along the positive $z$-axis after two successive total internal reflections at the conical surface.

We studied two cases of retro-reflection from conical dielectrics of differing refractive indices in order to understand the various aspects of SAM to OAM conversion. Figure 10 shows the characteristics of the reflected beam when a 90º glass cone of refractive index $n_{cone} = 1.55$, base radius = height = 6 µm, coated with a 100 nm-thick layer of refractive index $n_{coat} = 1.245$, is illuminated with the RCP Gaussian beam of Fig. 1 ($\lambda_o = 0.5$ µm, FWHM = 4 µm). For the chosen $n_{cone}$, the phase-difference between the s- and p-components of the light rays after each TIR is $\phi_s - \phi_p = 45°$, resulting in a net phase-shift (upon two reflections) of 90º and, therefore, complete conversion of the incident circular polarization to linear polarization. The reflected beam thus has no SAM, but is endowed with a certain amount of OAM.

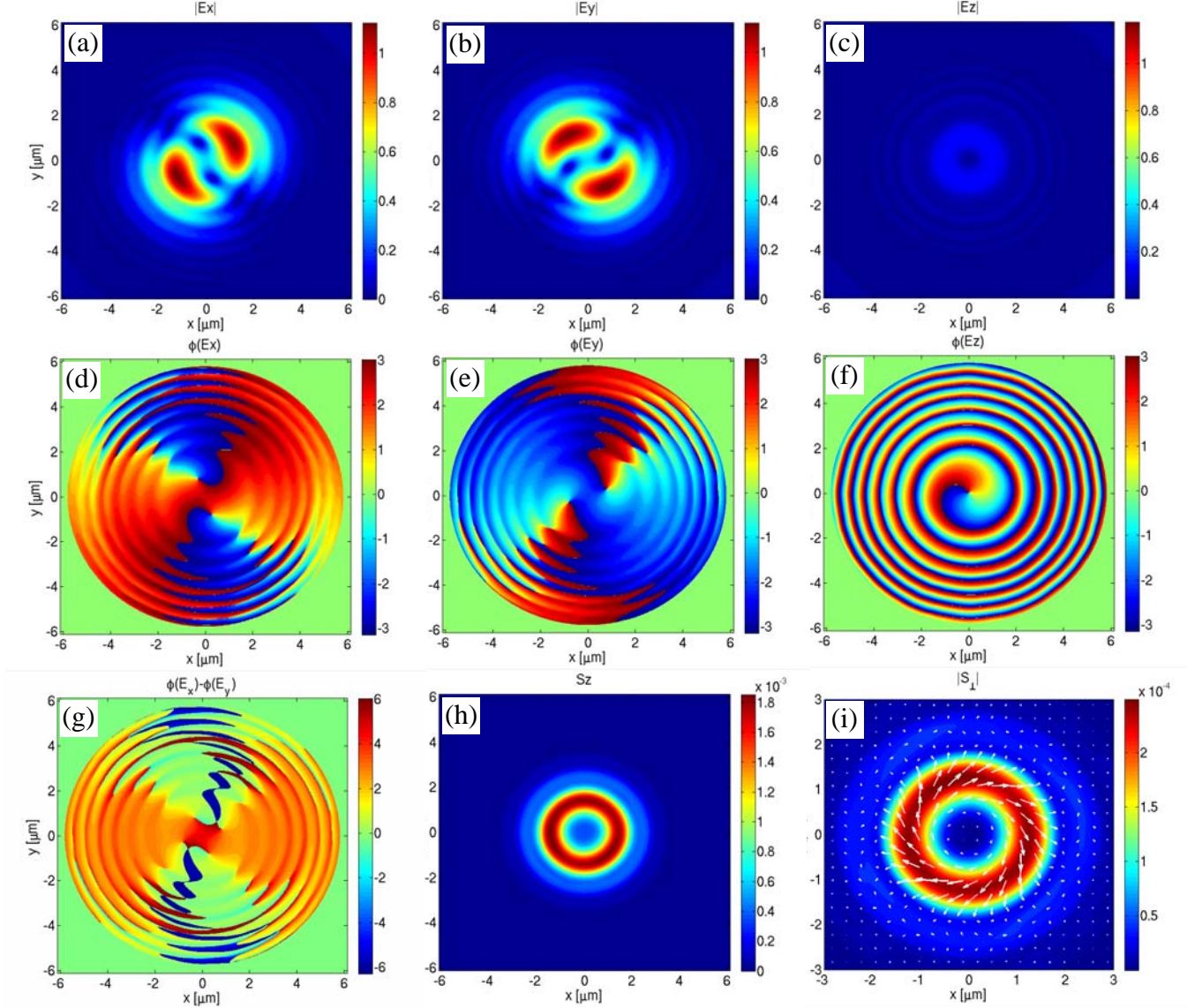

**Fig. 10** (Color online). Reflection of the RCP Gaussian beam of Fig. 1 from a glass cone ($n_{cone} = 1.55$, $R = 6$ µm, $h = 6$ µm, $n_{coat} = 1.245$, $t_{coat} = 100$ nm). For the chosen $n_{cone}$, the phase-difference imparted to the light's s- and p-components upon each TIR is 45°. The reflected beam is thus linearly polarized, as can be seen in the plot of $\phi(E_x) - \phi(E_y)$, which is nearly zero in some regions and almost 180° in others. Both $E_x$ and $E_y$ show $4\pi$ vorticity in their phase structure, but the corresponding OAM is only half as much as that of a full $4\pi$ vortex, because $|E_x|$ and $|E_y|$ are not uniformly distributed around $z$. The $z$-component of the field exhibits only a $2\pi$ vorticity. The plot of $S_z$ is doughnut-like, albeit with a partially-filled hole. The clockwise circulation of $\boldsymbol{S}_\perp$ around $z$ is almost entirely associated with the OAM of the (linearly-polarized) reflected beam. The proximity of the 45° angle of incidence on the conical surface to the critical TIR angle of 40.18° is responsible for a small fraction of the incident light leaking out of the cone. In our simulation, the actual fraction of the incident light that returned along $+z$ was 92.1%; the corresponding AM of the returning beam was 92.6% that of the incident.

When we computed the integrated Poynting vector in the cross-sectional plane of the reflected beam, we found it to contain only 92.1% of the incident optical power; this is due to the fact that, because of diffraction, a small fraction of the incident rays strike the conical surface at an incidence angle below the critical TIR angle; these rays are partially transmitted through the cone, a fact that could be readily confirmed by monitoring the transmitted optical power below the cone. We then proceeded to integrate the AM density $\boldsymbol{r} \times \boldsymbol{S}_\perp/c^2$ over the reflected beam's cross-section, and found it to be $\sim$92.6% of the incident AM. Thus, accounting for the leakage and aside from small numerical errors, the AM content of the reflected beam is seen to be the same as that of the incident beam. In other words, the incident SAM appears to have completely disappeared and replaced by an equal amount of OAM in the reflected beam. The net result is that no optical AM has been transferred (in the form of mechanical AM) to the glass cone.

We also confirmed that a lateral shift in the position of the cone relative to the beam center does not alter the main conclusions of the preceding analysis. When the cone was displaced by 2 µm along the $y$-axis, we found the general behavior of the spin and orbital angular momenta to remain the same. The total AM of the beam along the $z$-axis also retained its initial value (to within numerical errors), again verifying the impossibility of getting the cone to spin on its axis.

In our second set of simulations, we chose $n_\mathrm{cone} = 2.56$, and coated the top facet of the cone with a 78 nm-thick layer of refractive index $n_\mathrm{coat} = 1.6$. As before, the cone's base radius was equal to its height at 6 µm, and the incident RCP Gaussian beam had $\lambda_\mathrm{o} = 0.5$ µm and FWHM = 4 µm. The phase-shift introduced between the $s$- and $p$-components after each TIR is now $\phi_s - \phi_p = 79.3°$, resulting in a state of elliptical polarization upon retro-reflection from the cone. Figure 11 shows the characteristics of the reflected beam in this case. The reflected optical power is now close to 100% of the incident power (i.e., no leakage through the cone), and the reflected AM is also close to 100% that of the incident AM. While the incident AM is exclusively due to spin, the reflected beam contains a mixture of SAM and OAM. Once again, the reflected beam is seen to have automatically adjusted its OAM content in order to preserve the total AM of the incident beam, thus ensuring that the cone will not acquire any mechanical angular momentum. (The plots of $S_z$ and $\boldsymbol{S}_\perp$ in Fig. 11 show a certain departure from circular symmetry, which we believe is caused by insufficient numerical accuracy. The smallest pixel size $\Delta x = \Delta y = \Delta z = 5$ nm that we were able to use in these simulations produced a mesh that was apparently too coarse for the fields inside the high-index cone to properly converge to a circularly symmetric solution. In other respects, however, the simulation results depicted in Fig. 11 appear to be reliable.)

**8. Impossibility of AM transfer to transparent dielectrics of axial symmetry**. Nieminen *et al* have given a rigorous argument in support of the assertion that optical AM cannot be transferred to a transparent axisymmetric dielectric along its axis of symmetry [20]. In the present section, we outline a simple proof of this impossibility theorem in the special case where the *intensity* distribution throughout the object has axial symmetry around the axis of symmetry of the object.

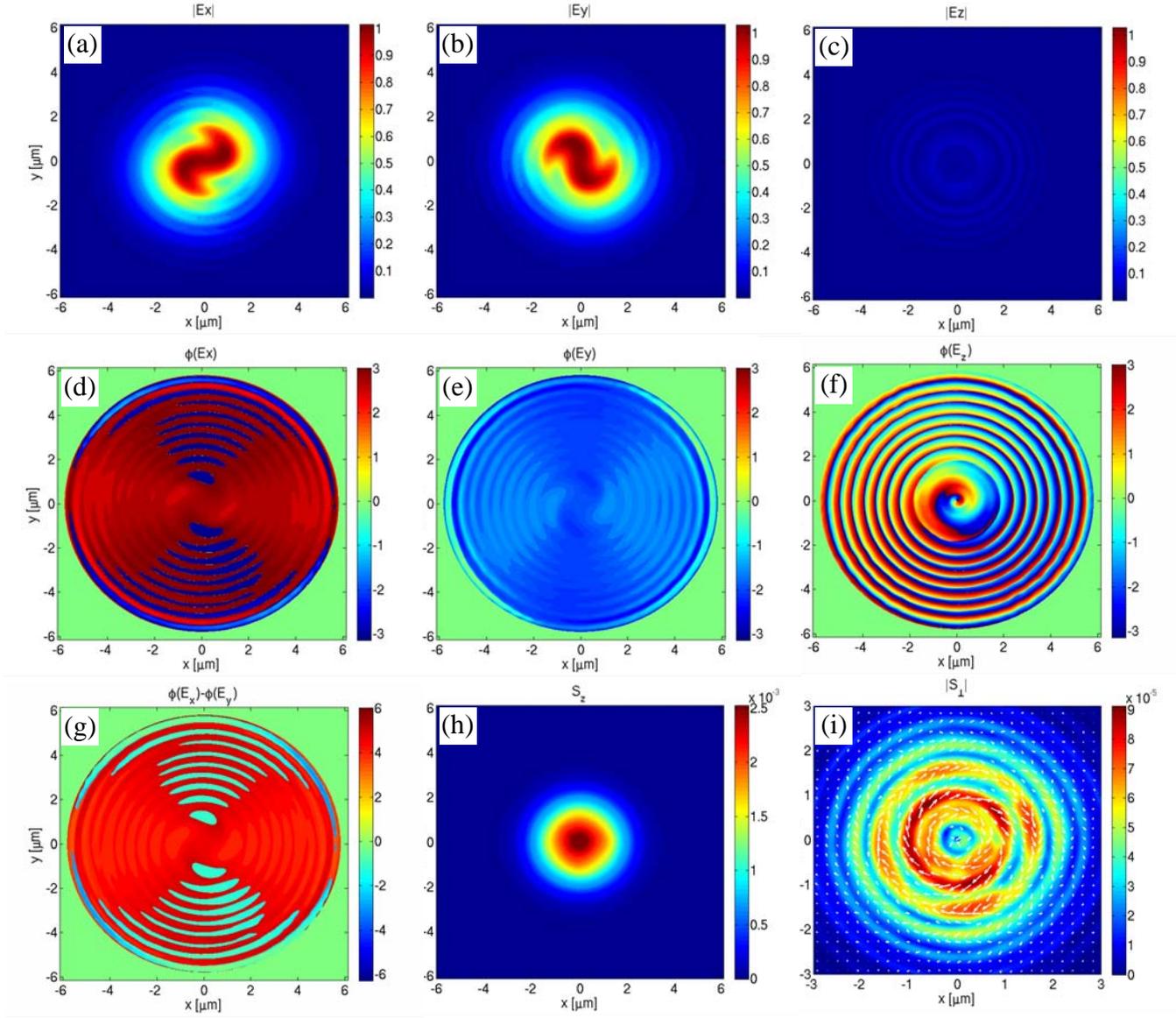

**Fig. 11** (Color online). Reflection of the RCP Gaussian beam of Fig. 1 from a glass cone ($n_{cone} = 2.56$, $R = 6$ μm, $h = 6$ μm, $n_{coat} = 1.6$, $t_{coat} = 78$ nm). For the chosen $n_{cone}$, the phase-difference imparted to the light's *s*- and *p*-components upon each TIR is 79.3°. The reflected beam is thus elliptically polarized with a fixed degree of ellipticity but varying orientations of the polarization ellipse. This may be seen in the plots of $|E_x|$, $|E_y|$, and $\phi(E_x) - \phi(E_y)$, with the latter showing as much as 43° phase variation over the beam's cross-section. Both $E_x$ and $E_y$ show slight variations in their phase profiles over the beam's cross-section, but no multiple-of-$2\pi$ vorticity, although $E_z$ exhibits a $2\pi$ vortex. The plots of $S_z$ and $\boldsymbol{S}_\perp$ no longer possess circular symmetry around *z*, which is probably an artifact caused by insufficient numerical accuracy. The clockwise circulation of $\boldsymbol{S}_\perp$ around *z* is characteristic of the beam's total AM, which contains a mixture of SAM and OAM. In our simulation, the fraction of the incident light that returned along the positive *z*-axis was nearly 100%; the corresponding AM of the returning beam was also about 100% that of the incident. The beam's AM is thus fully conserved upon reflection, although a fraction of its SAM has been converted to OAM.

Consider a transparent dielectric medium with axial symmetry around *z*. Let the incident beam similarly possess axial symmetry with respect to *z* in its various field amplitudes, although one must allow for the possibility of deviation from such symmetry in the form of optical vortices as well as circularly-polarized beams that exhibit a spiral phase around *z* in one or more field components. Even in the presence of such phase spirals, it is possible for the *total intensity* distribution ($|E_x|^2+|E_y|^2+|E_z|^2$) inside the transparent object to remain symmetric around *z*. (This was the case for the glass cone depicted in Fig. 10, and apparently *not* the case for that in Fig. 11.) The gradient of this intensity profile, which is proportional to the optical force-density exerted on the material medium, will thus have components along the radial (*r*) and vertical (*z*) directions of a cylindrical coordinate system, but no components will exist along the azimuthal (*ϕ*) direction. Such force distributions, of course, cannot produce any torques along *z*, as the force-density vectors everywhere remain co-planar with the *z*-axis. In addition, at the surface(s) of the dielectric object, there exist surface forces due to the fact that the perpendicular *E*-field component is generally discontinuous at such surfaces. These surface forces always being in the direction of the surface-normal, they cannot produce any torque along the *z*-axis. This completes the impossibility proof for the special case of an axially symmetric intensity distribution inside a transparent and axisymmetric dielectric object.

**9. Concluding remarks**. It is sometimes considered desirable to split the EM angular momentum into a spin part, associated with circular polarization, and an orbital part, which is intimately tied to the vorticity of the beam. The total AM of a light pulse with respect to an arbitrary point $r_o$ is the integral of $(r - r_o) \times S(r,t)/c^2$ over the spatial volume occupied by the pulse. This prescription applies whether the AM is due to the polarization state of the beam, or its vorticity, or a mixture of the two. In other words, one does not distinguish SAM from OAM when computing the total AM of an electromagnetic wave. Since the distinction cannot be based on an analysis of the Poynting vector profile, it must lie in the local or global properties of the *E*- and *H*-fields, and, perhaps more importantly, in the methods of monitoring such properties.

Suppose, for instance, that the field at and around a given point *r* is circularly polarized. If we place at *r* a small spherical particle of an absorptive material, the particle acquires some mechanical AM from the EM field and begins to rotate on its axis. The essential physics of this process involves the appearance (within the particle) of an induced dipole moment *p*, which co-rotates with the local *E*-field. The strength *p* of the dipole-moment is proportional to the local *E*-field, with the proportionality constant being the magnitude of the particle's electric susceptibility. The absorptive nature of the particle renders its susceptibility complex-valued. Absorption thus produces a lag between the induced dipole moment *p* and the local *E*-field, with the phase of the complex susceptibility determining the angle by which the rotating vector *p* lags behind the co-rotating *E*. The torque experienced by the particle is then given by *p* × *E*, which is responsible for the spinning of the particle on its axis. An isotropic and transparent particle would *not* have behaved in this way, because its induced dipole moment *p* would have been aligned with the *E*-field at all times. In contrast, a transparent birefringent particle *will* pick up some spin from the local field, as its birefringence produces the all-important angle between the induced dipole *p* and the local *E*-field. In all these examples, the local or global structure of the field's Poynting vector is irrelevant; what matters is that the EM field at point *r* has a net circular polarization, and that the experiment is designed to sense this local polarization state.

In general, interactions with material media can change not only the angular momentum content of a beam, but also its composition in terms of the relative abundance of spin and orbital momenta. This paper has shown, among other things, that simple retro-reflecting devices (in the form of hollow metallic or solid dielectric cones) can readily shift the balance of angular momentum in favor of SAM or OAM, without changing – in the absence of absorption – the overall AM content of a beam. A possible application of hollow metallic cones is in conjunction with broadband light such as white light or extremely short (femtosecond) light pulses. Unlike transmission-type phase-plates and liquid-crystal-based devices, metallic reflectors are largely free from chromatic aberrations, which makes them ideal for generating optical vortices across an entire input spectrum, or transforming a short pulse into a vortex without distorting the temporal profile of the pulse.

Axial symmetry plays an important role in transferring angular momentum from an EM wave to a material object. We saw in Sec. 5 that an axisymmetric PEC object cannot acquire any AM along its axis of symmetry, irrespective of the nature of illumination. Also, it was argued in Sec. 8 that a transparent object of axial symmetry cannot be made to spin around its axis, so long as the intensity of the EM field within the object remains axisymmetric with respect to that axis. We also pointed out the existence of a more general proof of the impossibility of AM transfer to transparent axisymmetric objects in the case of arbitrary illumination. Breaking the axial symmetry of the object thus becomes an important criterion if the goal is to impart angular momentum from an EM wave to the material object. The wedge-shaped reflector described in Sec. 3 is a good example of a simple device that lacks axial symmetry in its geometric structure, and has the ability to acquire AM from an incident beam. Another example is a slab of transparent birefringent crystal, such as a half-wave plate, which, by virtue of its crystal asymmetry (i.e., possessing different refractive indices along different crystallographic directions), can change the state of polarization of an incident light beam, say, from RCP to LCP, thereby picking up the change in the optical AM as the light passes through the slab.

This paper has focused exclusively on the properties of cones having an apex angle of 90º. The retro-reflecting property, however, is not limited to 90º cones; in fact, the property is shared among all hollow metallic cones with an apex angle of 180º/$m$, where $m$ is any positive integer. If $m$ happens to be even, each incident ray suffers $m$ reflections inside the cone before emerging on the opposite side while propagating anti-parallel to its direction of incidence. When $m$ is odd, each incident ray suffers ½($m-1$) reflections at oblique incidence, before landing on the conical surface at normal incidence. The ray then retraces its path and re-emerges at its point of entry while propagating in the reverse direction. We will explore the properties of such cones in a separate paper.

**Acknowledgement**. The authors are grateful to Brian Anderson, Poul Jessen, and Timo Nieminen for helpful discussions.

**References**

1. L. Allen, M. W. Beijersbergen, R. J. C. Spreeuw, and J. P. Woerdman, "Orbital angular-momentum of light and the transformation of Laguerre-Gaussian laser modes," Phys. Rev. A **45**, 8185-8189 (1992).
2. N. B. Simpson, K. Dholakia, L. Allen, and M. J. Padgett, "Mechanical equivalence of spin and orbital angular momentum of light: An optical spanner," Opt. Lett. **22**, 52-54 (1997).
3. L. Allen, M. J. Padgett, and M. Babiker, "The orbital angular momentum of light," Prog. Opt. **39**, 291-372 (1999).
4. J. H. Crichton and P. L. Marston, "The measurable distinction between the spin and orbital angular momenta of electromagnetic radiation," Electronic Journal of Differential Equations, Conference 04 (Mathematical Physics and Quantum Field Theory), pp. 37-50 (2000).
5. A. T. O'Neil, I. MacVicar, L. Allen, and M. J. Padgett, "Intrinsic and Extrinsic Nature of the Orbital Angular Momentum of a Light Beam," Phys. Rev. Lett. **88**, 053601 (2002).
6. S. M. Barnett, "Rotation of electromagnetic fields and the nature of optical angular momentum," J. Mod. Opt. **57**, 1339-1343 (2010).

7. M. E. J. Friese, H. Rubinsztein-Dunlop, J. Gold, P. Hagberg, and D. Hanstorp, "Optically driven micromachine elements," Appl. Phys. Lett. **78**, 547-549 (2001).
8. D. G. Grier, "A revolution in optical manipulation," Nature **24**, 810-816 (2003).
9. Y. Zhao, J. S. Edgar, G. D. M. Jeffries, D. McGloin, and D. T. Chiu, "Spin-to-Orbital Angular Momentum Conversion in a Strongly Focused Optical Beam," Phys. Rev. Lett. **99**, 073901 (2007).
10. T. A. Nieminen, A. B. Stilgoe, N. R. Heckenberg, and H. Rubinsztein-Dunlop, "Angular momentum of a strongly focused Gaussian beam," J. Opt. A: Pure Appl. Opt. **10**, 115005 (2008).
11. Y. Zhao, D. Shapiro, D. McGloin, D. T. Chiu, and S. Marchesini, "Direct observation of the transfer of orbital angular momentum to metal particles from a focused circularly polarized Gaussian beam," Optics Express **17**, 23316-22 (2009).
12. L. Marrucci, E. Karimi, S. Slussarenko, B. Piccirillo, E. Santamato, E. Nagali, and F. Sciarrino, "Spin-to-orbital conversion of the angular momentum of light and its classical and quantum applications," J. Opt. **13**, 064001 (2011).
13. K. I. Lee, J. A. Kim, H. R. Noh, and W. Jhe, "Single-beam atom trap in a pyramidal and conical hollow mirror," Opt. Lett. **21**, 1177-1179 (1996).
14. J. A. Kim, K. I. Lee, H. R. Noh, W. Jhe, and M. Ohtsu, "Atom trap in an axicon mirror," Opt. Lett. **22**, 117-119 (1997).
15. D. Fink, "Polarization effects of axicons," Applied Optics **18**, 581-582 (1979).
16. M. Endo, "Numerical simulation of an optical resonator for generation of a doughnut-like laser beam," Optics Express **12**, 1959-1965 (2004).
17. C. Konz and G. Benford, "Geometric absorption of electromagnetic angular momentum," Opt. Comm. **226**, 249-254 (2003).
18. T. A. Nieminen, "Comment on 'Geometrical absorption of electromagnetic angular momentum,' C. Konz, G. Benford," Opt. Comm. **235**, 227-229 (2004).
19. T. A. Nieminen, T. Asavei, V. L. Y. Loke, N. R. Heckenberg, and H. Rubinsztein-Dunlop, "Symmetry and the generation and measurement of optical torque," J. Quant. Spect. Rad. Trans. **110**, 1472-1482 (2009).
20. A. Taflove and S. C. Hagness, *Computational Electrodynamics: The Finite-Difference Time-Domain Method*, 2nd edition, Artech House, 2000.
21. A. E. Leanhardt, A. Görlitz, A. P. Chikkatur, D. Kielpinski, Y. Shin, D. E. Pritchard, and W. Ketterle, "Imprinting Vortices in a Bose-Einstein Condensate using Topological Phases," Phys. Rev. Lett. **89**, 190403 (2002).
22. M. V. Berry, "Quantal phase factors accompanying adiabatic changes," Proc. R. Soc. Lon. A **392**, 45-57 (1984).
23. M. V. Berry, "Interpreting the anholonomy of coiled light," Nature **326**, 277-278 (1987).
24. M. Kitano, T. Yabuzaki, and T. Ogawa, "Comment on 'Observation of Berry's Topological Phase by Use of an Optical Fiber,' " Phys. Rev. Lett. **58**, 523 (1987).
25. D. Akamatsu and M. Kozuma, "Coherent transfer of orbital angular momentum from an atomic system to a light field," Phys. Rev. A **67**, 023803 (2003).